\documentstyle[aps,preprint]{revtex}

\begin{document}

\newcommand{\ini}{\begin{equation}}
\newcommand{\fin}{\end{equation}}
\newcommand{\inir}{\begin{eqnarray}}
\newcommand{\finr}{\end{eqnarray}}
\newcommand{\inif}{\begin{figure}}
\newcommand{\finf}{\end{figure}}
\newcommand{\bc}{\begin{center}}
\newcommand{\ec}{\end{center}}
\def\ol{\overline}
\def\pa{\partial}
\def\ra{\rightarrow}
\def\ts{\times}
\def\df{\dotfill}
\def\bs{\backslash}
\def\dg{\dagger}

\preprint{DSF-13/2001}

\title{Fermion masses and mixings in gauge theories}

\author{D. Falcone}

\address{Dipartimento di Scienze Fisiche, Universit\`a di Napoli,
Via Cintia, Napoli, Italy}

\maketitle

\begin{abstract}
The recent evidence for neutrino oscillations stimulate us to discuss
again the problem of fermion masses and mixings in gauge theories. In the
standard model, several forms for quark mass matrices are equivalent. They 
become ansatze within most extensions of the standard model, where also 
relations between quark and lepton sectors may hold. In a seesaw
framework, these relations can constrain the scale of heavy neutrino mass,
which is often related to the scale of intermediate or unification gauge
symmetry. As a consequence, two main scenarios arise.
Hierarchies of masses and mixings may be explained by broken horizontal
symmetries.
\end{abstract}


\newpage

\tighten

\section{Introduction}

Gauge theories have become a standard framework for the description of fundamental
interactions \cite{quigg}. They rely on the invariance of the Lagrangian
with respect to a group of local transformations. If the vacuum state has a
restricted invariance, the theory is said to be spontaneously broken.
This condition allows the gauge bosons and the fermions to take a mass,
without spoiling the renormalizability.
Within gauge theories, fermion masses and mixings are important
physical parameters. Masses are intrinsic properties of particles, while
mixings are related to the possibility of their reciprocal transformations.
Currently, there are many experimental and theoretical developments
in this field. In particular, the SuperKamiokande Collaboration
has recently confirmed the oscillation of
atmospheric neutrinos as an explanation of the atmospheric neutrino
anomaly \cite{sk}, that is the deficit of neutrinos produced by cosmic
rays
in the atmosphere. There are also strong indications
of oscillation for solar neutrinos in order to explain the long standing
solar neutrino problem \cite{sun,bks}, the deficit of neutrinos produced by
nuclear reactions in the sun.

Neutrino oscillations are easily accounted for if
neutrinos have nonzero masses. However, in the
minimal standard model (MSM), that is the gauge theory for strong, weak
and electromagnetic interactions, based on the group $SU(3) \ts SU(2) \ts
U(1)$ and with only one complex Higgs doublet and no right-handed neutrinos,
such particles are massless.
Several extensions of the MSM allow for nonzero neutrino mass.
Some of them do not require the extension of the gauge group,
but only of the particle contents, for example adding
the right-handed neutrino or the Higgs triplet, although gauge
extensions such as the $SO(10)$ model,
which does include the right-handed neutrino in its fundamental
representation, provide a natural framework for
a small neutrino mass.
On the other hand, the study of neutrino masses and mixings should be related
to the more general problem of elementary fermion masses and mixings in gauge
and string theories, and in fact the SuperKamiokande results stimulate us to
consider again what we know about this interesting and difficult subject.

In this short review we deal with fermion mass terms and matrices in gauge theories,
taking into account the gauge allowed mass terms and the relation
between mass matrices and physical parameters (masses and mixings),
that is the quantities which are observable in principle.
Possible relations between quark and lepton masses are reported. Also
neutrino masses and lepton mixings are discussed.

In renormalizable gauge theories, fermion masses arise from
bare mass terms like $\ol{\psi} M_0 \psi$, and from Yukawa terms like
$\ol{\psi} Y \psi \phi$, which after spontaneous symmetry breaking (SSB)
give mass terms $\ol{\psi} M_1 \psi$, with $M_1=Yv$ and $v$ the vacuum
expectation value (VEV) of the Higgs field $\phi$. The structure of mass
terms is further determined by the fact that the components of the Higgs
representation which can have VEV different from zero must be neutral and
colorless, since charge and color are always conserved.
The gauge theories that we
consider here are the MSM plus three right-handed neutrinos (which we would like
to call the standard model (SM)), the left-right model (LRM)
$SU(3) \ts SU(2) \ts SU(2) \ts U(1)$, the Pati-Salam partial unification model
$SU(4) \ts SU(2) \ts SU(2)$, the $SU(5)$ and $SO(10)$ grand unification models. 
Our concern is mainly about the nonsupersymmetric theory. However,
some important comments on the supersymmetric version \cite{nilles} are included in the text.
We try to isolate the main features of each model, without entering
the detailed realizations existing in the literature.

A clear feature of the quark and charged lepton mass spectrum is the
hierarchy of masses belonging to different generations:
\ini
m_u \ll m_c \ll m_t,~~m_d \ll m_s \ll m_b,~~m_e \ll m_{\mu} \ll m_{\tau}.
\fin
We do not discuss in detail the origin of such hierarchies, and only mention
two mechanisms that have been proposed, namely the radiative
mechanism \cite{rad}, based on loop effects, and the
Froggatt-Nielsen mechanism \cite{fn}. 
The latter is based on broken horizontal (flavor or family) symmetries,\footnote{Horizontal
symmetries link together particles of different generations. On the contrary,
usual gauge symmetries are vertical, because they relate particles of the same
generation or family.}
generating effective mass operators through the coupling of ordinary fermions to
hypothetical heavy states (universal seesaw mechanism).
Nonsupersymmetric and supersymmetric scenarios which include the effects of
this mechanism are outlined in section VII.

\section{Standard model}

The standard description of strong, weak and electromagnetic
interactions is a gauge theory based on the group
$SU(3)_c \ts SU(2)_L \ts U(1)_Y$, with coupling constants $g_3$, $g_2$, $g_1$,
respectively.
The group $SU(3)_c$ describes the strong
interaction of quarks \cite{fgl}, while the product $SU(2)_L \ts U(1)_Y$
describes the electroweak model \cite{gws}, where a relationship between
weak and electromagnetic interactions stands out. The SSB by the Higgs
mechanism \cite{higgs} cuts the symmetry $SU(2)_L \ts U(1)_Y$ down to the
abelian symmetry $U(1)_Q$ of quantum electrodynamics.
In such a way the weak gauge bosons take a mass,
while gluons and the photon remain massless. For energy scales lower than the
electroweak symmetry breaking scale there are two exact symmetry, that is
color $SU(3)$ for the strong interaction and electric charge $U(1)$ for
the electromagnetic interaction. The SSB of the standard group requires the
existence of a neutral Higgs boson, not yet discovered. The
recent indication to the discovery, by LEP experiments, needs to be confirmed.
Let us summarize the particle contents of the MSM.

The classification of left-handed fermions belonging to the first generation
within
representations of the group $SU(3)_c \ts SU(2)_L \ts U(1)_Y$ is the following:
\ini
\left( \begin{array}{c}
       u \\ d
      \end{array} \right) \sim ({\bf 3},{\bf 2},1/6),~
      d^c \sim ( {\bf {\ol{3}}},{\bf 1},1/3),
     ~u^c \sim ( {\bf {\ol{3}}},{\bf 1},-2/3)
\fin
for quarks and antiquarks, and
\ini
\left( \begin{array}{c}
       \nu_e \\ e
      \end{array} \right) \sim ({\bf 1},{\bf 2},-1/2),~
      e^c \sim ({\bf 1},{\bf 1},1)
\fin
for leptons and the positron (the antineutrino is absent).
The twelve gauge bosons are the gluons
({\bf 8},{\bf 1},0), the $W$ bosons ({\bf 1},{\bf 3},0), and
the $B$ boson ({\bf 1},{\bf 1},0). In this writing the first number
indicates the representation of $SU(3)_c$, the second one the
representation of $SU(2)_L$ and the third
number is the hypercharge $Y$, with $Q=T_3+Y$, where $T_3$ is the quantum
number related to the third generator
of the group $SU(2)_L$ and $Q$ is the electric charge.\footnote{We denote
with a bar the conjugate representation. The product of a representation
with its conjugate gives the singlet. The adjoint representation, which
corresponds to gauge bosons, is autoconjugate or real.}
Right-handed charge-conjugate states belong to the conjugate representations
of (2),(3). Quarks are color triplets.
The other two generations have an analogous classification,
according to $u \sim c \sim t$, $d \sim s \sim b$, $e \sim \mu \sim \tau$.
There is also the scalar Higgs doublet $\varphi$,
\ini
\varphi=
\left( \begin{array}{c}
       \varphi^0 \\ \varphi^-
      \end{array} \right)
       \sim ({\bf 1},{\bf 2},-1/2),
\fin
which breaks $SU(2)_L \ts U(1)_Y$ down to $U(1)_Q$, due to its nonvanishing VEV.
Instead, as said above, $SU(3)_c$ is not broken.
This framework, in particular the presence of $SU(2)_L$ in the gauge group,
includes the V$-$A theory of weak interactions \cite{fgms}.
Moreover, the $SU(3)_c$ component allows for asymptotic freedom \cite{gwp}
and confinement \cite{wgw} of quarks and gluons inside hadrons.
  
The electroweak interactions of quarks and leptons are mediated by the weak
gauge bosons $W^+$, $W^-$ and $Z^0$, and by the photon $A^0$, which are
linear combinations of the bosons $W$ and $B$, that is
$Z^0=\cos \theta_W W^0 - \sin \theta_W B^0$,
$A^0=\sin \theta_W W^0 + \cos \theta_W B^0$,
where $\theta_W$ is the Weinberg angle, $\sin^2 \theta_W \simeq 0.23$.
Such interactions are summarized in the Lagrangian
${\cal L}_{int}={\cal L}_{CC}+{\cal L}_{NC}$, where
\ini
{\cal L}_{CC}=\frac{g_2}{\sqrt{2}}(J_{\mu}^+W^{-\mu}+J_{\mu}^-W^{+\mu})
\fin
describes charged current weak interaction and
\ini
{\cal L}_{NC}=\frac{g_2}{2\cos\theta_W}J_{\mu}^0Z^{\mu}+eJ_{\mu}^{em}A^{\mu}
\fin
describes neutral weak and electromagnetic interactions.
The parameter $g_2$ is the coupling constant related to the group $SU(2)_L$, 
the parameter $e$ is the coupling constant of electrodynamics, and
the electroweak relations $g_1 \cos\theta_W=g_2 \sin\theta_W=e$ hold.
Currents are defined as $J^-=(J^+)^{\dg}$,
\ini
J_{\mu}^+=\ol{u}\gamma_{\mu}d'+
\ol{c}\gamma_{\mu}s'+
\ol{t}\gamma_{\mu}b'
+\ol{\nu}_e\gamma_{\mu}e+
\ol{\nu}_{\mu}\gamma_{\mu}\mu+
\ol{\nu}_{\tau}\gamma_{\mu}\tau,
\fin
\ini
J_{\mu}^{em}=\sum_f Q_f\ol f\gamma_{\mu}f,
\fin
\ini
J_{\mu}^0=\sum_f \ol f\gamma_{\mu}(v_f-a_f\gamma_5)f,
\fin
with $v_f=T_{3f}-2Q_f \sin^2\theta_W$, $a_f=T_{3f}$, where $Q_f$ and $T_{3f}$ 
refer to the left-handed fermion field $f_L$.
In Eqns.(8),(9) both the left-handed and
the right-handed components are allowed, while in Eqn.(7) only left-handed
components appear, according to the V$-$A theory.

Primed fields in Eqn.(7) mean that the weak eigenstates $(d',s',b')$
are not equal to the
corresponding mass eigenstates $(d,s,b)$ but are linear combinations of them
\ini
\left( \begin{array}{c}
      d' \\ s' \\ b'
  \end{array} \right)=
 \left( \begin{array}{ccc}
 V_{ud} & V_{us} & V_{ub} \\
 V_{cd} & V_{cs} & V_{cb} \\
 V_{td} & V_{ts} & V_{tb} 
 \end{array} \right)
\left( \begin{array}{c}
      d \\ s \\ b
  \end{array} \right).
\fin
The unitary matrix $V$ connecting mass to weak (flavor) eigenstates is the
Cabibbo-Kobayashi-Maskawa (CKM) matrix \cite{c,km}.
The standard parametrization of this matrix is given by
\ini
V= \left( \begin{array}{ccc}
     c_{12}c_{13} & s_{12}c_{13} & s_{13}e^{-i\delta} \\
-s_{12}c_{23}-c_{12}s_{23}s_{13}e^{i\delta} &
c_{12}c_{23}-s_{12}s_{23}s_{13}e^{i\delta} &
s_{23}c_{13} \\
s_{12}s_{23}-c_{12}c_{23}s_{13}e^{i\delta} &
-c_{12}s_{23}-s_{12}c_{23}s_{13}e^{i\delta} &
c_{23}c_{13}
\end{array} \right)
\fin
where $c_{ij}=\cos\theta_{ij}$, $s_{ij}=\sin\theta_{ij}$ and $\delta$
is the CP violating phase.
A $3 \ts 3$ unitary matrix
contains three angles and six phases. Five phases in the CKM matrix can be
absorbed due to the freedom of five relative phases for the left-handed quarks.
The explicit form (11) is obtained by the product of three rotations in
sectors 2-3, 1-3, 1-2. Thus, three angles and one phase appear as
independent parameters.
It is well known that $s_{23}$ and $s_{13}$ are small, $O(10^{-2})$
and $O(10^{-3})$ respectively, hence $c_{23}\simeq c_{13}\simeq 1$
and we get
$s_{12} \simeq |V_{us}|$, $s_{23} \simeq |V_{cb}|$, $s_{13}=|V_{ub}|$.
Setting $\lambda=s_{12}\simeq 0.22$, $s_{23}=A\lambda^2$ and
$s_{13}e^{-i\delta}=A\lambda^3(\rho-i\eta)$, with $A,\rho,\eta$ of order 1,
we obtain the Wolfenstein parametrization \cite{wol} 
\ini
V \simeq \left( \begin{array}{ccc}
1-\frac{\lambda^2}{2} & \lambda & A\lambda^3 (\rho-i\eta) \vspace{0.2cm} \\ 
-\lambda & 1-\frac{\lambda^2}{2} & A\lambda^2  \vspace{0.2cm} \\    
A\lambda^3 (1-\rho-i\eta) & -A\lambda^2 & 1
   \end{array} \right)
\fin
where the four independent parameters are
$\lambda,~A,~\rho,~\eta$.
In the case of only two generations of quarks, the mixing matrix becomes a
rotation,
which depends on the unique parameter $\theta_C$, the Cabibbo angle \cite{c},
$\sin\theta_C = \lambda \simeq 0.22$, without the CP violating phase.
It was a great achievement by Kobayashi and Maskawa to show, before the
third generation was discovered, that if there were three generations of
quarks then the MSM allowed CP violation.

From Eqn.(12) we see that $V$ is
near the identity, that is quark mixings are small. Moreover, we have the
mixing hierarchy
$V_{ub} \ll V_{cb} \ll V_{us}$. These features probably have the same origin as
the mass hierarchies (1), as we will see.
Values of quark masses and mixings at several scales in the SM and in the
supersymmetric SM are collected in Ref.\cite{fk}.
Mass hierarchy of charged leptons is similar to that of down quarks, while
for up quarks it is enhanced: 
\ini
\frac{m_d}{m_s} \sim \frac{m_s}{m_b} \sim \lambda^2,~~
\frac{m_u}{m_c} \sim \frac{m_c}{m_t} \sim \lambda^4.
\fin
Any theory of fermion masses and mixings should explain these mass ratios
and the structure of Eqn.(12). As a matter of fact, we have only hints on
how such a pattern may arise.

\subsection{Fermion mass terms}

As we said in the introduction, fermion masses in gauge theories can be
generated by bare mass terms or Yukawa terms. These things must be
gauge invariant, they must transform as the gauge group singlet. Moreover,
a mass term is the Lorentz invariant part in the product of two
left-handed spinor fields of a particle type, but can also be written using the
bispinor notation (see the Appendix). We will use spinors in the product
of representations and bispinors in mass terms.
In the MSM it is not possible to build gauge invariant bare mass terms,
since no pair of multiplets in (2),(3) yields the singlet.
However, some Yukawa terms are allowed, because we have
\inir
({\bf 1},{\bf 2},-1/2)_f \ts ({\bf 1},{\bf 1},1)_f \ts
({\bf 1},{\bf 2},-1/2)_H &=& ({\bf 1},{\bf 1},0)+...
\\
({\bf 3},{\bf 2},1/6)_f \ts ({\bf \ol{3}},{\bf 1},1/3)_f
\ts ({\bf 1},{\bf 2},-1/2)_H &=& ({\bf 1},{\bf 1},0)+...
\\
({\bf 3},{\bf 2},1/6)_f \ts ({\bf {\ol{3}}},{\bf 1},
-2/3)_f \ts ({\bf 1},{\bf 2},1/2)_{\ol{H}} &=&
({\bf 1},{\bf 1},0)+...
\finr
where $H= \varphi$ and $\ol{H}= \text{i} \sigma_2 H^*$ are used (for
$SU(2)$ every representation is equivalent to its conjugate).
After the SSB they give the mass terms for charged leptons and quarks,
respectively
\ini
{\cal L}_m = \ol{e}_L M_e e_R+\ol{d}_L M_d d_R+\ol{u}_L M_u u_R,  
\fin
where $M_i=Y_i v$ are matrices in generation space,
$Y_i$ are Yukawa coupling matrices.
Sum with the hermitian conjugate is understood.
The physical parameter $v=174$ GeV is the VEV of the Higgs field $\varphi^0$
(the weak scale). At the tree level the masses of the weak gauge bosons are
$m_W=g_2 v/\sqrt2$ and $m_Z=m_W/\cos \theta_W$.
When we diagonalize the mass matrices $M_u$ and $M_d$, by means of biunitary
transformations, $V_{uL}^{\dg} M_u V_{uR}$, $V_{dL}^{\dg} M_d V_{dR}$, 
the CKM matrix $V_{CKM}=V_{uL}^{\dg} V_{dL}$ appears in the charged
current interaction. Unitary transformations with $V_{iL}$ and $V_{iR}$
diagonalize $M_i M_i^{\dg}$ and $M_i^{\dg} M_i$, respectively.

Some remarks are soon in order. First,
the value of $v$ is nearly equal to the top quark mass $m_t$. Moreover,
in the supersymmetric case two distinct Higgs doublets $H_1$
and $H_2$, with VEVs $v_1$ and $v_2$, are necessary.
One has $v_1^2+v_2^2=v^2$ and $v_2/v_1=\tan \beta$, so that $v_2=v \sin \beta$,
$v_1=v \cos \beta$. The doublet
$H_2$ generates $M_u$ while the doublet $H_1$ generates both $M_d$
and $M_e$, a possible hint towards the hierarchy $m_t \gg m_b \sim m_{\tau}$.
Considering only the third generation, in the MSM we have $m_t=y_t v$,
$m_b=y_b v$, so that $y_t \simeq 1$, $y_b \sim 10^{-2}$.
In the supersymmetric case
$m_t=y_t v \sin \beta$, $m_b=y_b v \cos \beta$, and $m_t \gg m_b$ may follow
from high $\tan \beta$, keeping $y_t \simeq y_b \simeq 1$ valid.
The last relation will be justified in the $SO(10)$
model.

In the MSM there is no mass term for neutrinos. However,
it is natural to complete the fermion spectrum by adding the analogue of $u^c$
in the lepton sector (see the classification (2),(3)), that is
the left-handed antineutrino 
$\nu^c \sim ({\bf 1},{\bf 1},0)$,
along with the right-handed neutrino.
Then, a bare Majorana mass term for the right-handed neutrino,
\ini
({\bf 1},{\bf 1},0)_f \ts ({\bf 1},{\bf 1},0)_f = ({\bf 1},{\bf 1},0),
\fin      
and a Yukawa term,
\ini
({\bf 1},{\bf 2},-1/2)_f \ts ({\bf 1},{\bf 1},0)_f \ts
({\bf 1},{\bf 2},1/2)_{\ol H}=({\bf 1},{\bf 1},0)+...,
\fin
providing a Dirac mass term,
are allowed. While the Dirac mass $m_D$ is expected to be of the same order of
magnitude as quark or charged lepton masses, because it is generated
by the term (19) which is very similar to terms (16) and
(14),(15), the Majorana mass $M_R$ is not constrained, because it is not
related to the SSB,
and can be very large. In such a case the full neutrino mass matrix
\ini
M=\left( \begin{array}{cc}
       0 & m_D \\
       m_D & M_R
     \end{array} \right)
\fin
gives a small eigenvalue $m_L \simeq m_D^2/M_R$ for the mass of a left-handed
Majorana neutrino and a large eigenvalue $m_R \simeq M_R$ for the mass of
a right-handed Majorana neutrino. This is the so-called seesaw
mechanism \cite{seesaw},
which can explain the smallness of the neutrino mass with respect to the
masses of charged fermions.\footnote{If only term (19) is introduced,
we have the problem
of why neutrino mass is so small with respect to the electron mass. The seesaw
mechanism is the most natural and elegant way to account for small neutrino mass,
although other mechanisms have been proposed. For example, a small neutrino mass
can be generated radiatively at one loop order \cite{zee} or at two loop order
\cite{babu}, introducing several new particles in the theory.
The possibility of a Yukawa coupling between the lepton doublet and a Higgs
triplet \cite{gr}, with small VEV, is now excluded by the LEP data on the
$Z^0$ width.}
A small mass for the left-handed neutrino is produced through the mixing with
a very heavy right-handed neutrino.
In the MSM the light neutrino mass term would appear as a
nonrenormalizable (dimension-5) effective operator which breaks lepton number
conservation by two units \cite{eoss}. For a numerical estimate on the value
of $M_R$, let us assume $m_D \sim m_t \sim 10^2$ GeV and
$m_L \simeq 5 \cdot 10^{-2}$ eV, the mass scale obtained by SuperKamiokande
\cite{sk}. Then we get $M_R \simeq 2 \cdot 10^{14}$
GeV. If one assumes $m_D \simeq m_{\tau} \simeq 1.8$ GeV, then
$M_R \simeq 6 \cdot 10^{10}$ GeV. In both cases a new (and very high) scale
appears in the theory.
For three generations one has a $6 \ts 6$ matrix
\ini
M= \left( \begin{array}{cc}
0 & M_D^T \\ M_D & M_R
\end{array} \right),
\fin
where $M_D$ and $M_R$ are $3 \ts 3$ matrices, and the eigenvalues of $M_R$
are much
bigger than the elements of $M_D$. The effective Majorana mass matrix for light
left-handed neutrinos is given by the seesaw formula
\ini
M_L \simeq -M_D^T M_R^{-1} M_D,
\fin
while for the heavy right-handed Majorana neutrinos it is $M_H \simeq M_R$.
The Majorana matrices $M_L$ and $M_R$ are symmetric, but the Dirac matrix
$M_D$ may be also asymmetric. After diagonalization of $M_e$ and $M_L$, a
lepton mixing matrix is induced in the charged weak interaction, allowing
for neutrino oscillations \cite{bgg}. In subsection IIC we discuss the scale and
structure of $M_R$ in the case of large lepton mixing and small Dirac mixing
(that is $M_D$ and $M_e$ almost diagonal).

\subsection{Quark mass matrices}

We are now interested in the Lagrangian formed by the quark mass terms
and charged weak interaction, simplified as
\ini
{\cal L}_M=\ol{u}_L M_u u_R+\ol{d}_L M_d d_R+g \ol{u}_L d_L W.
\fin
Diagonalizing $M_u$ and $M_d$ we get
\ini
{\cal L}_M=\ol{u}_L D_u u_R+\ol{d}_L D_d d_R+g \ol{u}_L V_{CKM} d_L W.
\fin
In this form of the Lagrangian we see a general property of masses and mixings.
Masses link a particle to itself, while mixings link a particle to another.
Apart from the weak coupling constant $g=g_2/\sqrt2$, Eqn.(24) contains
ten observables quantities, namely six quark masses, three angles and one phase,
while Eqn.(23) can have 36 real parameters in $M_u$ and $M_d$.
This count means that within the SM there is much freedom in the choice of quark
mass matrices.
In fact, without changing the ten observables in Eqn.(24), it is possible
to perform the following unitary transformations on quark fields in Eqn.(23):
\ini
u_L \ra U u_L,~d_L \ra U d_L,
\fin
\ini
u_R \ra V_u u_R,~d_R \ra V_d d_R.
\fin
Left-handed states transform together, while right-handed states $u_R$ and
$d_R$ are independent from each other. Of course, this fact is
related to the classification (2).
By using transformations (25),(26) one can reduce the number of independent
parameters in quark matrices till ten (minimal parameter basis).
For example, one can get two hermitian
mass matrices \cite{fj} (18 parameters) and in particular
one matrix diagonal and the other hermitian \cite{ma} (12 parameters).
In fact, $M_1$ can be made hermitian or diagonal by $U^{\dg} M_1 V_1$,
and $M_2$ hermitian by $U^{\dg} M_2 V_2$, since a polar decomposition theorem
allows one to write a matrix as product $HX$, where $H$ is hermitian
and $X$ unitary. For $M_u=D_u$
it is $M_d=V D_d V^{\dg}$, and for $M_d=D_d$ it is $M_u=V^{\dg} D_u V$.
Two phases in the hermitian mass matrix are fixed by the CKM representation
and we are left with ten observable parameters.
The numerical expression for $M_d$ when $M_u$ is diagonal looks like
\ini
M_d=\left( \begin{array}{ccc}
0.009 & 0.019 & 0.010 \\ 0.019 & 0.093 & 0.113 \\ 0.010 & 0.113 & 2.995
\end{array} \right),
\fin
where mass values are in GeV at the $M_Z$ scale. We note that
$M_{d33} \simeq m_b$, that is $Y_{d33} \simeq 1$ in the supersymmetric case
with large $\tan \beta$.

In addiction to the basis with $M_d$ hermitian, there are also several bases
with three zeros in $M_d$ \cite{f1,ft2}, so that the relation
$M_d M_d^{\dg}= V D_d^2 V^{\dg}$ allows one to get $M_d$.
We report here just two forms for it \cite{ft1,ft2}, with symmetric zeros,
\ini
M_d=\left( \begin{array}{ccc}
0 & 0.024 & 0 \\ 0.021 & 0.105 & 0.106 \\ 0 & 1.333 & 2.685
\end{array} \right),
\fin
\ini
M_d=\left( \begin{array}{ccc}
0 & 0 & 0.023 \\ 0 & 0.106 & 0.104 \\ 0.541 & 2.687 & 1.213
\end{array} \right).
\fin
There are three parameters (masses) in the diagonal $M_u$ and six real
mass parameters plus one phase, not written, in $M_d$.
In both asymmetric forms there are simple relations between elements.
In Eqn.(28) we have $M_{d12} \simeq M_{d21}$, $M_{d22} \simeq M_{d23}$ and
$M_{d33} \simeq 2 M_{d32}$. In Eqn.(29) $M_{d22} \simeq M_{d23}$ and
$M_{d32} \simeq 2 M_{d33}$. Moreover, Eqn.(28) can be written in the
approximate form
\ini
M_d=\left( \begin{array}{ccc}
0 & \sqrt{m_d m_s} & 0 \\ \sqrt{m_d m_s} & m_s & m_s \\
0 & \frac{m_b}{\sqrt5} & 2 \frac{m_b}{\sqrt5} 
\end{array} \right),
\fin
which is symmetric in the 1-2 sector and leads to the relations
\ini
V_{us} \simeq \sqrt{\frac{m_d}{m_s}},~
V_{cb} \simeq \frac{{3}}{\sqrt5} \frac{m_s}{m_b},~
V_{ub} \simeq \frac{1}{\sqrt5} \frac{\sqrt{m_d m_s}}{m_b}
\fin
between quark masses and weak mixings. The relation $M_{d22} \simeq M_{d23}$
is sensitive to the value of the phase $\delta$ in the CKM matrix, and holds
for $\delta=60^{\circ}-90^{\circ}$.

Another important (but not minimal) basis for quark mass matrices is
the nearest neighbour interaction (NNI) basis \cite{blm},
\ini
M_u=\left( \begin{array}{ccc}
0 & A' & 0 \\ A & 0 & B' \\ 0 & B & C 
\end{array} \right),~
M_d=\left( \begin{array}{ccc}
0 & D' & 0 \\ D & 0 & E' \\ 0 & E & F 
\end{array} \right).
\fin
The famous Fritzsch ansatz \cite{fri} is obtained if $M_u$ and $M_d$ are also
hermitian, in which case we have the approximate form
\ini
M_u=\left( \begin{array}{ccc}
0 & \sqrt{m_u m_c} & 0 \\
\sqrt{m_u m_c} & 0 & \sqrt{m_c m_t}  \\
0 & \sqrt{m_c m_t} & m_t 
\end{array} \right),~
M_d=\left( \begin{array}{ccc}
0 & \sqrt{m_d m_s} & 0 \\
\sqrt{m_d m_s} & 0 & \sqrt{m_s m_b}  \\
0 & \sqrt{m_s m_b} & m_b 
\end{array} \right).
\fin
This ansatz is now ruled out mainly by the huge value of the
top quark mass combined with the small value of $V_{cb}$. In fact the Fritzsch ansatz
implies two mass-mixing relations,
\ini
V_{us} \simeq \sqrt{\frac{m_d}{m_s}}-\sqrt{\frac{m_u}{m_c}},
\fin
\ini
V_{cb} \simeq \sqrt{\frac{m_s}{m_b}}-\sqrt{\frac{m_c}{m_t}}.
\fin
While the first relation is consistent with experiments, the second one
is ruled out for $V_{cb} \simeq 0.04$ and $m_t \simeq 180$ GeV.

Within the basis with $M_u$ and $M_d$ hermitian, it is always possible to
choose the mass matrices to have vanishing 1-3 and 3-1 elements \cite{fx}.
Then the further vanishing of the 1-1 element (but not also the 2-2) in
both matrices is still consistent with phenomenology \cite{cf,nmf},
and $M_u$, $M_d$ can be written in the approximate form
\ini
M_u=\left( \begin{array}{ccc}
0 & \sqrt{m_u m_c} & 0 \\
\sqrt{m_u m_c} & m_c & \sqrt{m_u m_t}  \\
0 & \sqrt{m_u m_t} & m_t 
\end{array} \right),~
M_d=\left( \begin{array}{ccc}
0 & \sqrt{m_d m_s} & 0 \\
\sqrt{m_d m_s} & m_s & \sqrt{m_d m_b}  \\
0 & \sqrt{m_d m_b} & m_b 
\end{array} \right),
\fin
yielding the relations (34) and
\ini
V_{cb} \simeq \sqrt{\frac{m_d}{m_b}}-\sqrt{\frac{m_u}{m_t}}.
\fin
Note that filling up position 2-2 in $M_d$ and $M_u$ we provide a direct
contribution
to the mass of quarks $s$ and $c$, thus allowing element 2-3, and hence
$V_{cb}$, to be lowered (compare Eqn.(33) with Eqn.(36)).

Hermitian minimal parameter bases (with no zeros in the diagonal) have been
classified in Ref.\cite{bbhl}. There are 18 such bases. They have one zero
in $M_u$ ($M_d$) and two in $M_d$ ($M_u$) (since these matrices are hermitian
we count two symmetric zeros as one). Phenomenologically viable
hermitian matrices with five and four zeros have been recently studied in
Refs.\cite{dv,kuo,rrr}.
For five zeros, only one form is actually viable \cite{cf,dv}. It can be
approximated as
\ini
M_u=\left( \begin{array}{ccc}
0 & 0 & \sqrt{m_u m_t} \\
0 & m_c & 0  \\
\sqrt{m_u m_t} & 0 & m_t 
\end{array} \right),~
M_d=\left( \begin{array}{ccc}
0 & \sqrt{m_d m_s} & 0 \\
\sqrt{m_d m_s} & m_s & \sqrt{m_d m_b}  \\
0 & \sqrt{m_d m_b} & m_b 
\end{array} \right),
\fin
yielding the relations
\ini
V_{us} \simeq \sqrt{\frac{m_d}{m_s}},~
V_{cb} \simeq \sqrt{\frac{m_d}{m_b}},~
V_{ub} \simeq \sqrt{\frac{m_u}{m_t}}.
\fin
It is important to notice that within the SM only the 1-1 zeros in (36)
and (38) are physically
meaningful, since the other zeros are basis zeros (see Ref.\cite{bbhl}).
All minimal parameter bases are equivalent in the SM.

The three matrix expressions (30) (with $M_u$ diagonal), (36) and (38),
in their different (approximate) form, have almost the same physical contents
on quark
masses and mixings in the SM, within the available experimental
informations.\footnote{Mixings in (33), (36) and (38) are given
by the formula $V_{ij} \simeq M_{uij}/M_{ujj}-M_{dij}/M_{djj}$, $j>i$.
In fact, for a symmetric $2 \ts 2$ mass matrix $M$
the mixing angle $\theta$ is given by
$\tan 2 \theta=2 M_{12}/(M_{22}-M_{11})$.
For a hierarchical matrix $\sin \theta \simeq M_{12}/M_{22}$. Moreover,
$V_{ij} \simeq V_{uij}-V_{dij}$. If $M$ is not symmetric
(for example in Eqn.(30)), the same calculation applies to $MM^{\dg}$.
The expression of mass-mixing relations depends on mass matrix model.}
In particular, matrices (38) and (30) are derived from two minimal parameter
bases.
Of course, matrices (38) give slightly different predictions for mixings,
with respect to matrices (36), but if one fills up position 1-1 with very
small entries, they become nearly equivalent. The approximate
vanishing of elements 1-1
is in correspondence with the approximate relations that we found for
asymmetric matrices.

In the hermitian cases the largest element
is by far the 3-3. Moreover, $M_{d33} \simeq m_b$ and $M_{u33} \simeq m_t$,
that is $Y_{d33} \simeq Y_{u33} \simeq 1$ in the supersymmetric model
with large $\tan \beta$ may hold.
This suggests that mixings and lighter generation masses are generated
on a different footing with respect to the third generation mass, which is
in some sense more fundamental.
This fact should be related to the Froggatt-Nielsen mechanism.
We see that the structure of matrices in Eqns.(36) and (38) allows for both
the hierarchies (1) and the CKM form (12). However, we stress again that many
other symmetric and asymmetric forms are possible. These forms become no more
equivalent only when the overall gauge symmetry is enlarged, as we will discuss
in the following sections.

\subsection{Neutrino masses and mixings}

Recent experiments show that neutrinos mix just like quarks do. However,
complications arise due to the possible addictional presence of the
Majorana mass term.
Let us consider the part of the SM Lagrangian related to lepton masses and
mixings,
\ini
{\cal L}=\ol{e_L} M_e e_R+\ol{{\nu}_L} M_D {\nu}_R+g \ol{{\nu}_L} e_L W+
\ol{({\nu}^c)_L} \frac{M_R}{2} \nu_R,
\fin
where $M_e$ is the mass matrix of charged leptons.
The effective Lagrangian in the seesaw mechanism is instead
\ini
{\cal L}_{ss}=\ol{e_L} M_e e_R+
\ol{{\nu}_L} \frac{M_L}{2} ({\nu}^c)_R+g \ol{{\nu}_L} e_L W+
\ol{({\nu}^c)_L} \frac{M_R}{2} \nu_R,
\fin
which shows the decoupling of heavy $\nu_R$ from light $\nu_L$.
Diagonalization of $M_e$ and $M_L$, with positive eigenvalues \cite{bz},
by means of (bi)unitary transformations $V_{eL}^{\dg} M_e V_{eR}$ and
$V_L^{\dg} M_L V_L^*$, respectively, and of $M_R$, gives
\ini
{\cal L}_{ss}=\ol{e_L} D_e e_R+
\ol{{\nu}_L} \frac{D_L}{2} ({\nu}^c)_R+g V_{MNS} \ol{{\nu}_L} e_L W+
\ol{({\nu}^c)_L} \frac{D_R}{2} \nu_R,
\fin
where $V_{MNS}=V_L^{\dg} V_{eL}$ is the (Maki-Nakagawa-Sakata)
lepton mixing matrix \cite{mns}. Contrary to the CKM matrix, neutrino oscillation experiments allow for
large and even maximal mixings in the MNS matrix.
It is easy to see that in Eqn.(41) we can go to a basis where $M_e$ is
diagonal, without changing masses and mixings appearing in Eqn.(42).
The relation between the weak (flavor) eigenstates $\nu_{\alpha}$
($\alpha=e,\mu,\tau$)
of Eqn.(41) and the mass eigenstates $\nu_i$
($i=1,2,3$) of Eqn.(42) is given by the MNS matrix,
$\nu_{\alpha}=V^{\dg}_{\alpha i} \nu_i = U_{\alpha i} \nu_i$,
and, if $M_e$ is diagonal, $M_L$ can be obtained from effective neutrino
masses and mixings, $M_L=U^{\dg} D_L U^*$,
with $D_L=$ diag $(m_1,m_2,m_3)$. The MNS matrix can be parametrized as the
standard form of the CKM matrix (Eqn.(11)) times a diagonal phase matrix
$D=\text{diag} (\text{e}^{\text{i} \alpha_1/2},\text{e}^{\text{i} \alpha_2/2},1)$.
Thus, two additional (Majorana) phases appear in the MNS matrix, due to the fact
that only the phases of charged leptons are free. Majorana phases do not
affect neutrino oscillations.

In order to determine the scale and structure of the mass matrix $M_R$,
we follow the (bottom-up) approach of Ref.\cite{f3}.
If there is no mixing, then
$M_L^{-1}=$ diag $(1/m_1,1/m_2,1/m_3)$. Instead, assuming a hierarchical
spectrum, $m_1 \ll m_2 \ll m_3$, for single maximal mixing
($U_{e2} =0,~U_{\mu 3}= 1/\sqrt2$, $U_{e3} \simeq 0$) one has 
\ini
M_L^{-1} \simeq \frac{1}{2} \left( \begin{array}{ccc}
\frac{2}{m_1} & 0 & 0 \vspace{0.2cm} \\ 
0 & \frac{1}{m_2} & -\frac{1}{m_2} \vspace{0.2cm} \\
0 & -\frac{1}{m_2} & \frac{1}{m_2}
\end{array} \right),
\fin
and for double maximal mixing
($U_{e2}= 1/\sqrt2,~U_{\mu 3}= 1/\sqrt2$, $U_{e3} \simeq 0$)
\ini
M_L^{-1} \simeq \frac{1}{2} \left( \begin{array}{ccc}
\frac{1}{m_1} & -\frac{1}{\sqrt2 m_1} & \frac{1}{\sqrt2 m_1} \vspace{0.2cm} \\
-\frac{1}{\sqrt2 m_1} & \frac{1}{2 m_1} & -\frac{1}{2 m_1} \vspace{0.2cm} \\
\frac{1}{\sqrt2 m_1} & -\frac{1}{2 m_1} & \frac{1}{2 m_1}
\end{array} \right).
\fin
Note that in matrix (44) all entries are of the same order.
This happens also for the 2-3 sector of matrix (43). In both cases the
leading form for $M_L$ was
\ini
M_L \sim \left( \begin{array}{ccc}
         0 & 0 & 0 \\ 0 & 1 & 1 \\ 0 & 1 & 1
\end{array} \right),
\fin
where a dominant block appears \cite{syv}.
Actually, in order to get the form of $M_R$ we have to assume something about
Dirac mass matrices. In many unified models (see for example
Refs.\cite{rrr,bky}),
the Dirac mixing is small, because lepton mass matrices are related to quark
mass matrices,\footnote{Note that, even in the supersymmetric SM,
$M_D$ and $M_u$ ($M_e$ and $M_d$)
are generated by the same Higgs doublet, $H_2$ ($H_1$), and a similarity,
$M_D \sim M_u$, $M_e \sim M_d$, can be assumed.} so that $M_e \simeq D_e$ and
\ini
M_D \simeq  \frac{m_{\tau}}{m_b} \text{diag} (m_u,m_c,m_t).
\fin
The leading form is $M_D \sim \text{diag}(0,0,1)$.
The factor $m_{\tau}/m_b$ takes into account the running of quark
masses with respect to lepton masses.
Also quark mixings run with the energy scale. However,
it has been shown \cite{bbo,br} that $V_{us}$ and $V_{ub}/V_{cb}$
are almost unchanged, while $V_{cb}$ and $V_{ub}$ may change at most
by 30 per cent, so that Dirac mixings remain small. For the present general
analysis we need not to perform a detailed calculation of running effects.
Then, from the inverted seesaw formula $M_R \simeq -M_D M_L^{-1} M_D^T$
we get the scales (the largest element in $M_R$)
\ini
M_{R33} \simeq \left( \frac{m_{\tau}}{m_b} \right)^2 \frac{m_t^2}{2m_2},
\fin
\ini
M_{R33} \simeq \left( \frac{m_{\tau}}{m_b} \right)^2 \frac{m_t^2}{4m_1},
\fin
for single and double maximal mixing, respectively.
From neutrino oscillation experiments we know that the mixing of atmospheric
($\nu_{\mu}$) neutrinos, related to $U_{\mu 3}$, is nearly maximal,
while the mixing of solar ($\nu_e$) neutrinos, related to $U_{e2}$, may be
small or large.
Therefore, from Eqns.(47),(48) we conclude that
for small mixing of solar neutrinos the scale
of $M_R$ depends on $1/m_2$, while for large mixing it depends on $1/m_1$.
We obtain
$M_R \sim 10^{15}$ GeV for the small mixing Mikheyev-Smirnov-Wolfenstein (MSW)
solution to the solar neutrino
problem,\footnote{For a recent account of the solar neutrino
problem and its oscillation solutions, see Ref.\cite{bks}.
MSW solutions refer to oscillations
in matter \cite{msw}.} $M_R \gtrsim 10^{16}$ GeV for large mixing MSW
(the favoured solution), 
$M_R \gtrsim 10^{17}$ GeV for (large mixing) low-$\Delta m^2$ MSW, and
$M_R \gtrsim 10^{18}$ GeV for (large mixing) vacuum oscillations.
The structure of the matrix $M_R$ is hierarchical and nearly diagonal,
with leading form
$M_R \sim \text{diag}(0,0,1)$, that is similar to $M_D$, but with an
enhanced hierarchy of masses and mixings \cite{pan}. Strong mass hierarchy
in the heavy neutrino mass matrix is indeed one of
the possible conditions in order to get a seesaw enhancement of lepton
mixing \cite{smir}. For a general classification of models with large lepton
mixing, see Ref.\cite{bado}.
In particular, for double large mixing we have found
$M_{Rij} \simeq m_{Di} m_{Dj}/m_1$. 
This last simple situation does not occur for the
inverse hierarchy $m_1 \simeq m_2 \gg m_3$, when however
$M_R \sim \text{diag}(0,0,1)$ again. The degenerate spectrum
$m_1 \simeq m_2 \simeq m_3$ is unnatural in the seesaw framework, if one
assumes a hierarchical $M_D$. In fact, in Eqn.(22), $M_R$ should work in such
a way as to cancel almost exactly the hierarchy of $M_D$.

At this stage, in the SM, the scale of $M_R$ is a new parameter of the theory.
However, in some extensions of the SM this scale is related to a SSB, and
thus, especially in the $SO(10)$ model, it is constrained. We will briefly
discuss this important issue at the end of section VI. Two distinct
theoretical patterns are related to it.

\section{Left-right model}

The SM is a successful theory. However, enlarged gauge symmetries have been
proposed, which are broken down to the SM symmetry at some high energy scale.
The simplest gauge extension of the SM involving a left-right analogy is
based on the gauge group $SU(3)_c \ts SU(2)_L \ts SU(2)_R \ts U(1)_{B-L}$
\cite{ps,lrm}. The groups $SU(2)_L$ and $SU(2)_R$ generate left-handed (V$-$A)
and right-handed (V+A) interactions, respectively, with coupling constants
$g_{2L}$ and $g_{2R}$. The V+A interactions are
suppressed by the high mass of $SU(2)_R$ gauge bosons. 

The classification of left-handed and right-handed
fermions is the following:
\ini
\left( \begin{array}{c}
       u \\ d
      \end{array} \right)_L \sim ({\bf 3},{\bf 2},{\bf 1},1/3),~
\left( \begin{array}{c}
       u \\ d
      \end{array} \right)_R \sim ({\bf 3},{\bf 1},{\bf 2},1/3)
\fin
\ini
\left( \begin{array}{c}
       \nu \\ e
      \end{array} \right)_L \sim ({\bf 1},{\bf 2},{\bf 1},-1),~
\left( \begin{array}{c}
       \nu \\ e
      \end{array} \right)_R \sim ({\bf 1},{\bf 1},{\bf 2},-1)
\fin
with $Q=T_{3L}+T_{3R}+(B-L)/2$. The generator $B-L$ is the difference
between baryon and lepton numbers. Charge-conjugate states belong to the
conjugate representations. Gauge bosons are
the gluons ({\bf 8},{\bf 1},{\bf 1},0),
the $W_L$ bosons ({\bf 1},{\bf 3},{\bf 1},0),
the $W_R$ bosons ({\bf 1},{\bf 1},{\bf 3},0), and the singlet
({\bf 1},{\bf 1},{\bf 1},0).
The Higgs fields needed to achieve the SSB and the
seesaw mechanism are $\varphi \sim$ ({\bf 1},{\bf 2},{\bf 2},0) and
$\Delta_R \sim$ ({\bf 1},{\bf 1},{\bf 3},2).
For example the bidoublet $\varphi$, written as
\ini
\varphi= \left( \begin{array}{cc}
\varphi_1^0 & \varphi_1^+ \\ \varphi_2^- & \varphi_2^0
\end{array} \right), 
\fin
breaks the SM group and gives Dirac masses
to quarks and leptons through the Yukawa terms
\inir
({\bf 3},{\bf 2},{\bf 1},1/3)_f \ts 
(\ol {\bf 3},{\bf 1},{\bf 2},-1/3)_{\ol{f}} \ts
({\bf 1},{\bf 2},{\bf 2},0)_{H,\ol{H}} & = & ({\bf 1},{\bf 1},{\bf 1},0)+... \\
({\bf 1},{\bf 2},{\bf 1},-1)_f \ts 
({\bf 1},{\bf 1},{\bf 2},1)_{\ol{f}} \ts
({\bf 1},{\bf 2},{\bf 2},0)_{H,\ol{H}} & = & ({\bf 1},{\bf 1},{\bf 1},0)+...,
\finr
with $H=\varphi$ and $\ol{H}=\sigma_2 H^* \sigma_2$. This happens because of
its VEV $\varphi_1^0=k_1$, $\varphi_2^0=k_2$. As a consequence,
the Dirac neutrino mass is generated on the same footing as the other fermion
masses, in particular the charged lepton mass.
Four Yukawa terms yield Dirac masses, as in the SM, but with a different pattern.
Quark mass matrices are written as $M_u=r k_1+s k_2^*$, $M_d=r k_2+s k_1^*$,
where $r$, $s$ are Yukawa coupling matrices.  
The triplet $\Delta_R$ breaks the left-right model to the SM
and gives a Majorana mass
to the right-handed neutrino through the Yukawa term
\ini
({\bf 1},{\bf 1},{\bf 2},-1)_f \ts 
({\bf 1},{\bf 1},{\bf 2},-1)_{f} \ts
({\bf 1},{\bf 1},{\bf 3},2)_H=({\bf 1},{\bf 1},{\bf 1},0)+....
\fin
If $k_{1,2}$ and $v_R$ are the VEVs of $\varphi_{1,2}^0$ and $\Delta_R$,
respectively, the full neutrino mass matrix is in the form
\ini
M \sim 
\left( \begin{array}{cc}
       0 & k_{1,2} \\
       k_{1,2} & v_R
     \end{array} \right)
\fin
and, since $k_{1,2} \ll v_R$, the seesaw mechanism holds.
All fermion masses are generated by the SSB and the mass of the
right-handed neutrino is related to the scale of left-right breaking \cite{ms}.
This is in contrast with the SM, where the right-handed neutrino mass
is produced by a bare mass term, unless a singlet Higgs field is included
on purpose.

In LRMs, if $k_1=k_2$, the Dirac neutrino masses are expected to be
similar to charged lepton masses
(and the up quark masses similar to the down quark masses, up-down symmetry)
and as a consequence the scale for $M_R$
is lowered by three or four orders with respect to the case (46)
(quark-lepton symmetry). However, up quark masses must be different from down
quark masses, so that in a minimal model $k_1 \ne k_2$ is required.
In general, mass matrices are arbitrary. If $k_2=0$ we have $r$ and $s$ very
different from each other.

\subsection{Right-handed mixings}

Let us discuss how the inclusion of $SU(2)_R$ in the gauge group allows one
to select, in principle, viable forms for quark mass matrices among the
several SM bases. The key input would be some knowledge of right-handed
interactions.
For the quark mass and charged weak current terms we have the Lagrangian
\ini
{\cal L}_M=\ol{u}_L M_u u_R+\ol{d}_L M_d d_R+
g_L \ol{u}_L d_L W_L+g_R \ol{u}_R d_R W_R
\fin
and after diagonalization of quark mass matrices
\ini
{\cal L}_M=\ol{u}_L D_u u_R+\ol{d}_L D_d d_R+g_L \ol{u}_L V_L d_L W_L+
g_R \ol{u}_R V_R d_R W_R,
\fin
with $V_L=V_{CKM}$, $V_R=V_{uR}^{\dg} V_{dR}$.
Right-handed currents appear on the same footing of the left-handed currents,
except for the fact that $m_{W_R} \gg m_{W_L}$. The mixing matrix $V_R$
contains three angles and six phases, since the phases of right-handed
quarks are fixed because they must cancel with the left-handed phases
in diagonalized mass matrices.
The transformations that do not change the observables in Eqn.(57) are
the following:
\ini
u_L \ra U u_L,~d_L \ra U d_L,
\fin
\ini
u_R \ra V u_R,~d_R \ra V d_R.
\fin
Left-handed states and also right-handed states transform together,
according to the classification (49).
Therefore, it is possible to have $M_1$ diagonal by $U^{\dg} M_1 V$ but
$M_2$ is fixed to become $U^{\dg} M_2 V$ and cannot be hermitian
or with three zeros in general as it happens in the SM. The bases
considered within the SM become strong ansatze in the LRM, giving the same
$V_L$ but different $V_R$.
Thus, a true test of mass matrices
should be done in a left-right extension of the SM.
In fact, a direct way to study quark mass matrices is to look at right-handed
mixings \cite{achi}.
For example, matrix (28) gives small right-handed mixings, while matrix
(29) gives a large $V_{cb}^R$.
A systematic analysis of quark mass matrices with $M_u$ diagonal and $M_d$
containing three zeros, within LRMs, has been performed in Ref.\cite{ft2}.
Right-handed mixings are obtained through the relation
$M_d^{\dg} M_d=V_R D_d^2 V_R^{\dg}$.
It is worth noting that for the hermitian models considered in subsection IIB
the right-handed mixings are equal to the left-handed ones.
This condition occurs in the case of manifest left-right symmetry
\cite{lansan}, when $g_{2L}=g_{2R}$ above the left-right
scale and $k_1$, $k_2$ are real.
In that case, also a Higgs triplet $\Delta_L$, with VEV $v_L \ll k_{1,2}$,
has to be introduced, which gives a direct contribution to the mass of
left-handed neutrinos \cite{ms2}.

By using transformations (58),(59) the form of both $M_u$ and $M_d$ may be
changed. However, we need other observable parameters, for example new mixings,
to select viable forms
for such models of mass matrices. These new physical parameters exist in
some extensions of the LRM, notably the Pati-Salam model (section IV)
and the $SO(10)$ model (section VI).

\section{Partial unification}

Lepton number is considered as the fourth color in the Pati-Salam model,
based on the gauge group $SU(4)_{C} \ts SU(2)_L \ts SU(2)_R$ \cite{ps}. 
This model preserves the left-right gauge analogy of LRM and in addiction the
color $SU(3)$ symmetry is extended to $SU(4)$ in order to include lepton number.
Fermions belong to the ({\bf 4},{\bf 2},{\bf 1}) and ({\bf 4},{\bf 1},{\bf 2})
and their conjugate representations,
\ini
\left( \begin{array}{cc}
u & \nu \\ d & e
\end{array} \right)_L \sim ({\bf 4},{\bf 2},{\bf 1}),~
\left( \begin{array}{cc}
u & \nu \\ d & e
\end{array} \right)_R \sim ({\bf 4},{\bf 1},{\bf 2}),
\fin
and gauge bosons to ({\bf 15},{\bf 1},{\bf 1}),
({\bf 1},{\bf 3},{\bf 1}), ({\bf 1},{\bf 1},{\bf 3}).
Left-handed states belong to one multiplet.
New interactions between quarks and leptons, and hence new mixings, arise.
In analogy with the previous model, the minimal Higgs contents
is ({\bf 1},{\bf 2},{\bf 2}), $(\ol{\bf 10},{\bf 1},{\bf 3})$, giving Dirac
masses by the two Yukawa terms
\ini
({\bf 4},{\bf 2},{\bf 1})_f \ts (\ol {\bf 4},{\bf 1},{\bf 2})_{\ol f} \ts
({\bf 1},{\bf 2},{\bf 2})_{H,\ol{H}}=({\bf 1},{\bf 1},{\bf 1})+...
\fin
and a Majorana mass to the right-handed neutrino by
\ini
({\bf 4},{\bf 1},{\bf 2})_f \ts ({\bf 4},{\bf 1},{\bf 2})_{f} \ts
(\ol{\bf 10},{\bf 1},{\bf 3})_{H}=({\bf 1},{\bf 1},{\bf 1})+....
\fin
Due to the fact that quarks and leptons
belong to the same multiplets, there is no more the freedom in fermion
mass matrices. The different forms for quark mass matrices considered above
correspond now to physically distinguished mechanisms,
even for the hermitian cases. For example, $V_{ub}$
has a direct component in model (38), while has an indirect origin in
model (36). If we use transformations (58),(59), some mixing between quarks
and leptons will change. Also the zeros appearing in SM bases are physically
meaningful.

Moreover, term (61), which involves the $SU(4)$ singlet, 
gives the mass relations
\ini
M_e=M_d,~M_{\nu}=M_u
\fin
(from now on we set $M_{\nu} \equiv M_D$), while analogous terms with
$({\bf 15},{\bf 2},{\bf 2})_{H,\ol{H}}$
would give
\ini
M_e=-3M_d,~M_{\nu}=-3M_u,
\fin
due to the form
$\text{diag} (1,1,1,-3)$ for the VEV of the adjoint {\bf 15}. Therefore,
the $SU(4)$ symmetry leads to simple relations between quark and
lepton mass matrices,
that is to quark-lepton symmetry. But if we combine the effect of the singlet
and the adjoint, simple relations are lost in general.
Quarks and leptons are unified in the Pati-Salam model, but in general not
the coupling
constants. This is achieved in the $SO(10)$ model. Before, we discuss the
prototype of unification models, that is the $SU(5)$ model.

\section{SU(5)}

Grand unification assumes that the gauge group describing the fundamental
interactions is simple and that it is broken down to the SM group
in one or more steps, each of them related to a mass scale and a residual
symmetry group. Therefore, the evolution of coupling constants must converge
to one point at the unification scale.
The simplest grand unification theory (GUT) is based on the unitary group
$SU(5)$ \cite{gg}, broken in one step down to the SM group.
The minimal nonsupersymmetric $SU(5)$ model does not agree with experimental
data, since the three coupling constants $g_3$, $g_2$, $g_1$ do not meet
at a single point
\cite{no}, proton lifetime is predicted too short and $\sin^2 \theta_W$ too small.
However, the supersymmetric version \cite {susy5} is reliable \cite{si},
with a predicted scale for superpartners, $M_S$, around $10^3$ GeV.
Proton lifetime results to be $1 \ts 10^{35 \pm 1}$ yr,
and the current experimental lower limit by SuperKamiokande is $3.3 \ts 10^{33}$ yr \cite{lim}.

Left-handed fermions belong to the $\ol{\bf 5}$ and {\bf 10} representations,
because under the group $SU(3) \ts SU(2) \ts U(1)$ these representations are
decomposed as
\inir
\ol {{\bf 5}} &=& (\ol {{\bf 3}},{\bf 1},1/3)+({\bf 1},{\bf 2},-1/2)
\\
{\bf 10} &=& ({\bf 3},{\bf 2},1/6)+(\ol {{\bf 3}},{\bf 1},-2/3)+
({\bf 1},{\bf 1},1),
\finr
in such a way that $d^c$, $e$, $\nu$ belong to the $\ol{\bf 5}$ and
$u$, $d$, $u^c$, $e^c$ to the {\bf 10}.
Gauge fields belong to the adjoint {\bf 24},
\ini
{\bf 24}=({\bf 8},{\bf 1},0)+({\bf 1},{\bf 3},0)+({\bf 1},{\bf 1},0)+
({\bf 3},{\bf 2},-5/6)+(\ol {{\bf 3}},{\bf 2},5/6),
\fin
whose terms refer to gluons, electroweak bosons and lepto-quarks.
The breaking of $SU(5)$ down to the SM group is obtained by a {\bf 24}
of Higgs fields and the further breaking to $SU(3) \ts U(1)$ by a ${\bf 5}$.

In the simplest case, fermion masses are generated by the Yukawa coupling with
the {\bf 5} of Higgs, while the {\bf 24} does not contribute. Since
\inir
\ol{\bf 5} \ts \ol{\bf 5} &=& \ol{\bf 10}+\ol{\bf 15} \\
\ol{\bf 5} \ts {\bf 10} &=& {\bf 5}+{\bf 45} \\
{\bf 10} \ts {\bf 10} &=& \ol{\bf 5}+\ol{\bf 45}+\ol{\bf 50},
\finr
we have two main terms
\inir
\ol{\bf 5}_f \ts {\bf 10}_f \ts  \ol{\bf 5}_H & = & {\bf 1}+... \\
{\bf 10}_f \ts  {\bf 10}_f \ts  {\bf 5}_{\ol H} & = & {\bf 1}+...,
\finr
which give mass to down quarks and charged leptons, and up quarks, respectively,
with the relation
\ini
M_e=M_d^T,
\fin
while $M_u$ is independent.
Also the inclusion of the {\bf 45} of Higgs fields can contribute to fermion
masses by the Yukawa terms
\inir
\ol{\bf 5}_f \ts  {\bf 10}_f \ts  \ol{\bf 45}_H & = & {\bf 1}+... \\
{\bf 10}_f \ts  {\bf 10}_f \ts  {\bf 45}_{\ol H} & = & {\bf 1}+...,
\finr
yielding the relation
\ini
M_e=-3M_d^T.
\fin
In the minimal $SU(5)$, just like in the MSM, the neutrino is massless.
However, if a
singlet $(\nu^c)_L$ is introduced, a Dirac mass is produced by the Yukawa
term
\ini
\ol{\bf 5}_f \ts {\bf 1}_f \ts {\bf 5}_{\ol H} ={\bf 1}+...,
\fin
and a Majorana mass for the right-handed neutrino
by the bare term ${\bf 1}_f \ts {\bf 1}_f$, or, if also a singlet Higgs is
introduced, by the Yukawa term ${\bf 1}_f \ts {\bf 1}_f \ts {\bf 1}_H$. 
Then a seesaw mechanism can work. One can also have a direct contribution to
the mass of left-handed neutrinos through the Yukawa term
${\bf \ol{5}}_f \ts {\bf \ol{5}}_f \ts {\bf 15}_H$.
Note that, since $M_u$ is generated by a
different term with respect to $M_{d,e}$, one can expect, as is the case,
the up quark mass hierarchy to be different from the down quark and charged
lepton mass hierarchies, which are similar to each other. This effect has
some analogy with that we have seen in supersymmetric SM, but is enforced by the
quadratic presence of ${\bf 10}_f$ in Eqn.(72).

The mass matrix relation (73) gives $m_d=m_e$, $m_s=m_{\mu}$,
$m_b=m_{\tau}$
at the unification scale and $m_d/m_e=m_s/m_{\mu}=m_b/m_{\tau} \simeq 3$ at
the low scale \cite{begn}. Thus only $m_b=m_{\tau}$ at the unification scale
is reliable. However, appropriate couplings with both $\ol{\bf 5}_H$ and
$\ol{\bf 45}_H$ give the suggestive Georgi-Jarlskog ansatz \cite{gj}
\ini
M_u=\left( \begin{array}{ccc}
          0 & A & 0 \\ A & 0 & B \\ 0 & B & C
          \end{array} \right),~
M_d=\left( \begin{array}{ccc}
          0 & D & 0 \\ D & E & 0 \\ 0 & 0 & F
          \end{array} \right),~
M_e=\left( \begin{array}{ccc}
          0 & D & 0 \\ D & -3E & 0 \\ 0 & 0 & F
          \end{array} \right),
\fin
where the relation $m_b=m_{\tau}$ is retained, and
$m_{\mu} \simeq 3 m_s$, $m_d \simeq 3 m_e$. Hence, at the low scale, the good
relations $m_b \simeq 3 m_{\tau}$, $m_s \simeq m_{\mu}$, $m_d \simeq 9 m_e$
are obtained. The matrix $M_u$ has the Fritzsch form.
The agreement with data is much better in the supersymmetric case
\cite{dhr}. However, a more confident choice for the mass matrices would be, for example,
that of
filling up element 2-2 in $M_u$ and element 2-3 in $M_{d,e}$, in order to have
the viable form with mass matrices having similar structure,
considered in subsection IIB. Yukawa unification $y_b=y_{\tau}$,
coming out from Eqn.(74), is reliable.
The transpose in Eqn.(73) or (76) can nicely be used to obtain large
lepton mixing retaining small quark mixing, if one takes asymmetric
matrices. Some models that use this feature, in the NNI basis,
are in Refs.\cite{ho,br}.

Since $u_R$ and $d_R$ belong to different multiplets, in $SU(5)$
we still have a freedom in quark mass matrices similar to the SM case, so
that without loss of generality $M_u$ can be taken diagonal (and $M_d$ with
three zeros). This is not true in the $SO(10)$ model.
Moreover, while in the $SU(4) \ts SU(2) \ts SU(2)$ model both $M_e$ and
$M_{\nu}$ are related to $M_d$ and $M_u$, respectively, in the $SU(5)$ model
$M_e$ is related to $M_d$, but $M_u$ and $M_{\nu}$ are independent
from each other. By using the Georgi-Jarlskog trick and the quark mass
matrix (30), one is led to propose \cite{ft3}
\ini
M_e= \frac{m_{\tau}}{m_b} \left( \begin{array}{ccc}
0 & \sqrt{m_d m_s} & 0 \\ \sqrt{m_d m_s} & -3 m_s & \frac{m_b}{\sqrt5} \\
0 & m_s & 2 \frac{m_b}{\sqrt5} 
\end{array} \right),
\fin
while $M_{\nu}$ could be diagonal, for example
$M_{\nu} \sim \text{diag} (m_u,m_c,m_t)
\sim \text{diag} (\lambda^8,\lambda^4,1) m_t$, or
$M_{\nu} \sim \text{diag} (m_d,m_s,m_b) \sim \text{diag} (m_e,m_{\mu},m_{\tau}) 
\sim \text{diag} (\lambda^4,\lambda^2,1)m_b$, or
$M_{\nu} \sim \text{diag} (\lambda^2,\lambda,1)m_{b,t}$.
A possible way to select the form of $M_{\nu}$ is by means of its impact within
the baryogenesis via leptogenesis mechanism \cite{ft3} (see also Ref.\cite{by}),
which is based on the decay of heavy neutrinos \cite{fyl}.
The $SU(5)$ model retains more freedom for mass matrices with respect to the
$SO(10)$ model that we are going to discuss.

\section{SO(10)}

As we said in the previous section, the nonsupersymmetric $SU(5)$ model
is not supported by experiment,
while supersymmetric $SU(5)$ is reliable. Another alternative to the
simplest $SU(5)$ is the $SO(10)$ model \cite{10}, which has also some advantages.
The orthogonal group $SO(10)$ includes both $SU(5)$ and
$SU(4) \ts SU(2) \ts SU(2)$.
Under $SU(5)$, the fundamental spinorial {\bf 16} representation of $SO(10)$
is decomposed as
\ini
{\bf 16}=\ol{\bf 5}+{\bf 10}+{\bf 1},
\fin
that is all left-handed fermions,
including the charge-conjugate states, belong to a single representation
and there is a natural place for $(\nu^c)_L$ in the $SU(5)$ singlet.
This fact makes the $SO(10)$ model more compelling, and possibly more
predictive for fermion masses, with respect to the $SU(5)$ model.
Gauge bosons are in the adjoint
${\bf 45}={\bf 24}+{\bf 1}+{\bf 10}+\ol{\bf 10}$,
where {\bf 24} are the $SU(5)$ bosons.
Under $SU(4) \ts SU(2) \ts SU(2)$ the {\bf 16} is decomposed as
\ini
{\bf 16}= ({\bf 4},{\bf 2},{\bf 1})+(\ol{{\bf 4}},{\bf 1},{\bf 2}).
\fin
The breaking to the SM
can occur through one, two, or more steps.
In a typical breaking chain \cite{dkp} the $SO(10)$ symmetry is broken down to
$SU(4) \ts SU(2) \ts SU(2)$ at the unification scale
$M_U \sim 10^{16}$ GeV by a ${\bf 210}_H$, then to the SM
at the intermediate scale $M_I \sim 10^{11}$ GeV by a ${\bf 126}_H$,
and the final
breaking is due to a ${\bf 10}_H$. Only the SM (or $SU(5)$) singlet in
${\bf 126}_H$, which corresponds to $({\bf 10},{\bf 1},{\bf 3})$ in the
Pati-Salam group, takes the VEV.

Looking at the product
${\bf 16} \ts {\bf 16}={\bf 10}+{\bf 126}+{\bf 120}$, with
${\bf 10}+{\bf 126}$ the symmetric part and {\bf 120} the antisymmetric
part, we realize that,
since representations {\bf 10} and {\bf 120} are real,
and the ${\bf 126}$ is complex,
the Yukawa terms which can give mass to the fermions are
\inir
{\bf 16}_f \ts {\bf 16}_f \ts  {\bf 10}_H &=& {\bf 1}+... \\
{\bf 16}_f \ts {\bf 16}_f \ts \ol{\bf 126}_H &=& {\bf 1}+... \\
{\bf 16}_f \ts {\bf 16}_f \ts {\bf 120}_H &=& {\bf 1}+....
\finr
However, we can use also two {\bf 10}s or two {\bf 120}s.
In such a case we obtain from term (82) the relations (63),
from term (83) the relations (64), and no relation from term (84).
These relations come out because of the $({\bf 1},{\bf 2},{\bf 2})$ component
of ${\bf 10}$, and the $({\bf 15},{\bf 2},{\bf 2})$
component of $\ol{\bf 126}$. Since the ${\bf 120}$ contains both components,
it yields no relation in the general case. The situation is similar to the
Pati-Salam model.
If one {\bf 10} is used, then in the nonsupersymmetric case we have
$|M_e|=|M_{\nu}|=|M_u|=|M_d|$, relations not viable. 
In general, the {\bf 10} and $\ol{\bf 126}$ contributions are symmetric and the
${\bf 120}$ contribution is antisymmetric.
A Majorana mass for the right-handed neutrino is obtained from the
coupling with the $SU(5)$ singlet contained in $\ol{\bf 126}$, related to the
scale of intermediate symmetry breaking.
The simplest seesaw mechanism is then in the form
\ini
M \sim 
\left( \begin{array}{cc}
       0 & {\bf 10} \\
       {\bf 10} & \ol{\bf 126}
     \end{array} \right),
\fin
but ${\bf 120}_H$ and $\ol{\bf 126}_H$ can contribute to Dirac masses and
$\ol{\bf 126}_H$ can have a component contributing to light neutrino mass.
The Georgi-Jarlskog ansatz is obtained by means of ${\bf 10}_H$ and
$\ol{\bf 126}_H$ \cite{hrr,rrr}, with mass matrices necessarily symmetric.
In this case the Dirac neutrino mass matrix is
related to the up quark mass matrix, in such a way that the calculation of
subsection IIC applies. Yukawa unification $y_b=y_{\tau}$,
coming out from Eqn.(82), is
reliable both in the nonsupersymmetric case, when it is realized at the
intermediate scale, and in the supersymmetric case, when it holds at the
unification scale.
The further unification $y_t=y_b=y_{\tau}$ is reliable only for
large $v_2/v_1$ or $\tan \beta$ \cite{lp}. In any case, Eqn.(82) gives also
$y_t=y_{\nu}$.
Therefore, a minimal model of fermion masses
includes only {\bf 10}s and {\bf 126}s. Two {\bf 10}s and one {\bf 126} provide
Dirac masses and another {\bf 126} gives a mass to right-handed neutrinos.
For a study of right-handed mixings in $SO(10)$, when asymmetric matrices
and {\bf 120}s are used, again in the NNI form, see Ref.\cite{am}.

We notice that in nonsupersymmetric $SO(10)$ it is $M_R \sim M_I$,
while as seen in subsection IIC the calculation gives $M_R \gtrsim M_U$.
There are two main ways to solve
this problem. First, if the matrix $M_R$ has a roughly off-diagonal form
and $M_{R33} \sim 0$
\cite{f3,off}, which can be obtained by some suitable cancellations
among effective neutrino parameters.
Second, in supersymmetric $SO(10)$, which can be broken in one step
to the SM, so that $M_R \sim M_U$. We notice also that for
vacuum oscillations the scale of $M_R$, as calculated in subsection
IIC, is well above the unification scale. This solution is not favoured from the
theoretical point of view. On the other hand the MSW solutions are consistent
with the unification scale. The large mixing MSW solution is also favoured by
data fits \cite{fits}.

\section{Outlook}

Understanding the pattern of fermion masses and mixings is a key subject
in modern particle physics. However, despite many efforts,
few advances have been obtained.
The values of quark masses and mixings are sufficiently well known,
but we do not have a confirmed theory about the origin of their spectrum.
On the other hand,
neutrino masses and lepton mixings are a new field of research both for
experiment and for theory. In extensions of the SM, the lepton sector is
often linked to the quark sector. Nevertheless, new observables can appear.
To date we have no experimental evidence about them.
Roughly speaking, the LRM leads to up-down symmetry, and the Pati-Salam
model to quark-lepton symmetry. The $SO(10)$ model preserves the features of
the Pati-Salam model, while the $SU(5)$ model only relates $M_d$ to $M_e$.

An indication towards a theory for fermion masses and
mixings could be obtained through the union of the Froggatt-Nielsen
mechanism with the Georgi-Jarlskog ansatz,
namely hierarchical entries in
mass matrices (see for example Eqn.(36)) together with simple relations
between the quark and
lepton sectors. In this case the zeros appearing in mass matrices are
approximate and in fact high powers of some small parameter \cite{ir}. Then,
a mass matrix is roughly in the form
\ini
M_{ij} \sim \epsilon^{q(f_i)+q(\ol{f}_j)}
\fin
where $q$ are charges related to the horizontal symmetry,
and the breaking parameter $\epsilon$ should be related to $\lambda$.
Usually $\epsilon$ is the ratio between the VEV of a heavy scalar field,
which breaks the horizontal symmetry, and the mass of very heavy fermions.
Several choices for this symmetry are possible.
For example, abelian $U(1)$ or nonabelian $U(2)$.
For some general considerations on broken horizontal symmetries, see
Ref.\cite{lns}.
The explicit realization of a compelling model can be much involved
\cite{ilr}.
Nevertheless, through this paper
we are led to two main general scenarios for fermion mass and mixing
parameters, preferably within the $SO(10)$ model.

In the nonsupersymmetric scenario the hierarchy $m_t \gg m_b$ is due to a double
Higgs contribution, the relation $m_b \simeq m_{\tau}$ holds at the
intermediate scale so that the values of $m_b$ and $m_{\tau}$ at the low
scale are understood as a running effect, Dirac matrices are nearly diagonal,
$M_R$ has a nearly off-diagonal form
at the intermediate scale.

In the supersymmetric scenario $m_t \gg m_b$ because $\tan \beta$ is
large,\footnote{The requirement of a large value for $\tan \beta$ has nontrivial
consequences for supersymmetric phenomenology \cite{beta}.}
the relation $m_b \simeq m_{\tau}$ holds at the unification scale running
at the low scale, Dirac matrices are nearly diagonal, also
$M_R$ has a nearly diagonal form at the unification scale.
The last remark makes the supersymmetric scenario more homogeneous.

For an exact horizontal symmetry, only the third generation would be massive,
and mixing angles vanish. Symmetry breaking terms gradually appear in mass
matrices as powers of $\epsilon$ or $\lambda$, generating the hierarchy
of masses and mixings, according to what we have seen in previous
sections. Different sets of matrix entries arise at different orders in
horizontal symmetry breaking and details depend on the assignment of horizontal
charges. Since the heavy fields responsible for this mechanism can
have masses above the unification scale, we see that a full account of
fermion masses and mixings is likely to be
realized towards the Planck scale ($M_P \sim 10^{19}$ GeV),
where quantum gravity effects in particle
physics become important. The leading theoretical framework for describing
this kind of physics is string theory \cite{string}, which indeed includes
horizontal symmetries.
Therefore, as a conclusion, we can say that the subject of fermion masses
and mixings has many phenomenological and theoretical implications.

\appendix
\section*{Dirac and Majorana mass terms}

In this appendix we write the Dirac and Majorana mass terms using both the
bispinor and the (Weyl) spinor notations. Let us consider a Dirac bispinor
$$
\varphi=\left( \begin{array}{c}
\psi_L \\ 0 \end{array} \right)+
\left( \begin{array}{c}
0 \\ \psi_R \end{array} \right)=
\varphi_L+\varphi_R,
$$
where the spinors $\psi_L$ and $\psi_R$ are the left-handed and
right-handed chirality
components. With this field $\varphi$ we can build the invariant term
$$
\ol{\varphi}\varphi=\ol{\varphi_R}\varphi_L+\ol{\varphi_L}\varphi_R,
$$
which corresponds to the Dirac mass term. The conjugate bispinor is
$$
\varphi^c=\left( \begin{array}{c}
(\psi^c)_L \\ 0 \end{array} \right)+
\left( \begin{array}{c}
0 \\ (\psi^c)_R \end{array} \right)=
\left( \begin{array}{c}
\sigma_2 \psi_R^* \\ 0 \end{array} \right)+
\left( \begin{array}{c}
0 \\ -\sigma_2 \psi_L^* \end{array} \right)=
(\varphi^c)_L+(\varphi^c)_R,
$$      
where $\sigma_2$ is a Pauli matrix, and we have
$\varphi^c=\gamma_2 \varphi^*$, so that 
$\ol{\varphi^c} \varphi^c=\ol{\varphi}\varphi$.
Let us define also (autoconjugate) Majorana bispinors:
$$
\varphi_l=\left( \begin{array}{c}
\psi_L \\ 0 \end{array} \right)+
\left( \begin{array}{c}
0 \\ -\sigma_2 \psi_L^* \end{array} \right)=
\left( \begin{array}{c}
\psi_L \\ 0 \end{array} \right)+
\left( \begin{array}{c}
0 \\ (\psi^c)_R \end{array} \right)=
\varphi_L+(\varphi^c)_R,
$$      
$$
\varphi_r=\left( \begin{array}{c}
\sigma_2 \psi_R^* \\ 0 \end{array} \right)+
\left( \begin{array}{c}
0 \\ \psi_R \end{array} \right)=
\left( \begin{array}{c}
(\psi^c)_L \\ 0 \end{array} \right)+
\left( \begin{array}{c}
0 \\ \psi_R \end{array} \right)=
(\varphi^c)_L+\varphi_R.
$$      
The left-handed (right-handed) field $\varphi_l$ ($\varphi_r$) can be expressed
in terms of the left-handed (right-handed) component $\psi_L$ ($\psi_R$) only.
There is an invariant term
$$
\ol{\varphi_l}\varphi_l=\ol{(\varphi^c)_R}\varphi_L+\ol{\varphi_L}(\varphi^c)_R,
$$
which corresponds to the left-handed Majorana mass term,
and another invariant term
$$
\ol{\varphi_r}\varphi_r=\ol{\varphi_R}(\varphi^c)_L+
\ol{(\varphi^c)_L}\varphi_R,
$$
which corresponds to the right-handed Majorana mass term.
Since $(\varphi_R)^c=(\varphi^c)_L$ and $(\varphi_L)^c=(\varphi^c)_R$ we have also
$
\ol{\varphi_l}\varphi_l=\ol{(\varphi_L)^c}\varphi_L+\ol{\varphi_L}(\varphi_L)^c
$
and
$
\ol{\varphi_r}\varphi_r=\ol{\varphi_R}(\varphi_R)^c+\ol{(\varphi_R)^c}\varphi_R.
$
In every mass term written above, the second term is the
hermitian conjugate (h.c.) of the first one. Note that a left-handed bispinor
is always coupled to a right-handed bispinor.

Previous mass terms are written using bispinors, but they can be expressed also
by means of spinors.
In fact, the Dirac invariant can be also written as
$$
(\psi^c)_L^T \sigma_2 \psi_L + h.c.~~ \text{or}~~
\psi_L^T \sigma_2 (\psi^c)_L + h.c.,
$$
and the two Majorana invariants as
$$
\psi_L^T \sigma_2 \psi_L + h.c.
$$
for the left-handed neutrino and
$$
(\psi^c)_L^T \sigma_2 (\psi^c)_L + h.c..
$$
for the right-handed neutrino. Remember that in $SU(2) \ts SU(2)$,
which acts as a representation of the Lorentz group, we have
$({\bf 2},{\bf 1}) \ts ({\bf 2},{\bf 1})=({\bf 1},{\bf 1})+({\bf 3},{\bf 1})$
and every mass term corresponds to the singlet $({\bf 1},{\bf 1})$.
Here, a left-handed spinor is coupled to another left-handed spinor.
Notice also that a factor 1/2 must be included in the Majorana terms,
due to the autoconjugation property of Majorana fields.
A Dirac mass term conserves lepton number and electric charge, while a Majorana
mass term violates lepton number by two units and is allowed only for neutral
particles. For example, the neutrino may be either of the Dirac or of the
Majorana type. If it is a Majorana particle, the neutrinoless double
beta decay can occur. Therefore, the discovery of this decay would prove
the Majorana nature for neutrinos.
If both the Dirac and the Majorana mass terms are present,
mass eigenstates are of the Majorana type.

\end{document}